\documentclass[sigconf,natbib,screen]{acmart}

\usepackage{natbib}
\usepackage{algorithm}
\usepackage{algorithmicx}
\usepackage{algpseudocode}
\usepackage{amsmath}
\usepackage{array}
\usepackage{booktabs}
\usepackage[normalem]{ulem}
\usepackage{footmisc}
\usepackage{graphicx}
\usepackage{epstopdf}
\usepackage{multirow}
\usepackage[inline]{enumitem}
\usepackage[skip=0pt]{caption}
\usepackage{subfigure}
\usepackage{xcolor}
\usepackage{pifont}
\usepackage{acronym}

\copyrightyear{2020} 
\acmYear{2020} 
\setcopyright{acmcopyright}\acmConference[SIGIR '20]{Proceedings of the 43rd International ACM SIGIR Conference on Research and Development in Information Retrieval}{July 25--30, 2020}{Virtual Event, China}
\acmBooktitle{Proceedings of the 43rd International ACM SIGIR Conference on Research and Development in Information Retrieval (SIGIR '20), July 25--30, 2020, Virtual Event, China}
\acmPrice{15.00}
\acmDOI{10.1145/3397271.3401081}
\acmISBN{978-1-4503-8016-4/20/07}

\AtBeginDocument{%
  \providecommand\BibTeX{{%
    \normalfont B\kern-0.5em{\scshape i\kern-0.25em b}\kern-0.8em\TeX}}}

\acrodef{MF}{matrix factorization}
\acrodef{RP}{rating prediction}
\acrodef{CM}{collaborative memory}
\acrodef{MR}{meta recommender}
\acrodef{CF}{collaborative filtering}
\acrodef{FCF}{federated collaborative filtering}
\acrodef{RG}{rise-dimensional generation}
\acrodef{MetaMF}{meta matrix factorization}
\acrodef{SVD}{singular value decomposition}
\acrodef{PCA}{principal component analysis}
\acrodef{PMF}{probabilistic matrix factorization}
\acrodef{NMF}{non-negative matrix factorization}
\acrodef{RBM}{restricted Boltzmann machine}
\acrodef{AE}{autoencoder}
\acrodef{MLP}{multi-layer perceptron}
\acrodef{CNN}{convolutional neural network}
\acrodef{NCF}{neural collaborative filtering}
\acrodef{DMF}{deep matrix factorization}
\acrodef{MAE}{Mean Absolute Error}
\acrodef{MSE}{Mean Square Error}

\newcommand{\squeeze}{\vspace*{-2mm}}

\author{Yujie Lin}
\orcid{}
\affiliation{%
\institution{Shandong University}
\city{Qingdao}
\country{China}
}
\email{yu.jie.lin@outlook.com}

\author{Pengjie Ren}
\orcid{}
\affiliation{%
\institution{University of Amsterdam}
\city{Amsterdam}
\country{The Netherlands}
}
\email{p.ren@uva.nl}
\authornote{Co-corresponding author.}

\author{Zhumin Chen}
\orcid{}
\affiliation{%
\institution{Shandong University}
\city{Qingdao}
\country{China}
}
\email{chenzhumin@sdu.edu.cn}

\author{Zhaochun Ren}
\orcid{}
\affiliation{%
\institution{Shandong University}
\city{Qingdao}
\country{China}
}
\email{zhaochun.ren@sdu.edu.cn}

\author{Dongxiao Yu}
\orcid{}
\affiliation{%
\institution{Shandong University}
\city{Qingdao}
\country{China}
}
\email{dxyu@sdu.edu.cn}
\authornotemark[1]

\author{Jun Ma}
\orcid{}
\affiliation{%
\institution{Shandong University}
\city{Qingdao}
\country{China}
}
\email{majun@sdu.edu.cn}

\author{Maarten de Rijke}
\orcid{0000-0002-1086-0202}
\affiliation{%
\institution{\mbox{}\hspace*{-4mm}\mbox{University of Amsterdam \& Ahold Delhaize}}
\city{Amsterdam}
\country{The Netherlands}
}
\email{m.derijke@uva.nl}

\author{Xiuzhen Cheng}
\orcid{}
\affiliation{%
\institution{Shandong University}
\city{Qingdao}
\country{China}
}
\email{xzcheng@sdu.edu.cn}

\renewcommand{\shortauthors}{Lin et al.}

\acmSubmissionID{319}

\begin{document}
\fancyhead{}

\title{Meta Matrix Factorization for Federated Rating Predictions}

\begin{abstract}
Federated recommender systems have distinct advantages in terms of privacy protection over traditional recommender systems that are centralized at a data center.
With the widespread use and the growing computing power of mobile devices, it is becoming increasingly feasible to store and process data locally on the devices and to train recommender models in a federated manner.
However, previous work on federated recommender systems does not fully account for the limitations in terms of storage, RAM, energy and communication bandwidth in a mobile environment.
The scales of the models proposed are too large to be easily run on mobile devices.
Also, existing federated recommender systems need to fine-tune recommendation models on each device, which makes it hard to effectively exploit \acl{CF} information among users/devices.

Our goal in this paper is to design a novel federated learning framework for \ac{RP} for mobile environments that operates on par with state-of-the-art fully centralized \ac{RP} methods.
To this end, we introduce a federated \ac{MF} framework, named \ac{MetaMF}, that is able to generate private item embeddings and \ac{RP} models with a meta network.
Given a user, we first obtain a collaborative vector by collecting useful information with a \acl{CM} module.
Then, we employ a \acl{MR} module to generate private item embeddings and a \ac{RP} model based on the collaborative vector in the server.
To address the challenge of generating a large number of high-dimensional item embeddings, we devise a \acl{RG} strategy that first generates a low-dimensional item embedding matrix and a rise-dimensional matrix, and then multiply them to obtain high-dimensional embeddings.
We use the generated model to produce private \acp{RP} for the given user on her device.

\ac{MetaMF} shows a high capacity even with a small \ac{RP} model, which can adapt to the limitations of a mobile environment.
We conduct extensive experiments on four benchmark datasets to compare \ac{MetaMF} with existing \ac{MF} methods and find that \ac{MetaMF} can achieve competitive performance.
Moreover, we find \ac{MetaMF} achieves higher \ac{RP} performance over existing federated methods by better exploiting \acl{CF} among users/devices.
\end{abstract}



\maketitle

\acresetall


\section{Introduction}
Traditionally, recommender systems are organized in a centralized fashion, i.e., service providers hold all data and models at a data center.
As even an anonymized centralized dataset still puts user privacy at risk via combinations with other datasets~\citep{sweeney-2000-uniqueness}, federated or decentralized recommender systems are increasingly being considered so to realize privacy-aware recommendations~\cite{ammad2019federated,barbosa2018distributed,DBLP:journals/corr/WangLCZQH15}.
In federated recommender systems, a global model in the server can be trained from user-specific local models on multiple mobile devices (e.g., phones, laptops, smartwatches), ensuring that users' interaction data never leaves their devices. 
Such recommender systems are capable of reducing the risk of leaking private user data.

Larger recommendation models need more space for storage, more RAM for running programs, more energy for calculation, and more communication bandwidth for downloading or updating.
Unlike fully centralized recommender systems at a data center, federated recommender systems that need to run on local devices have stricter requirements on the scale of the model.
Previous work on federated recommender systems~\citep{ziegler2004semantic,massa2007trust,DBLP:journals/corr/WangLCZQH15,ammad2019federated,chen2018federated} neglects to fully account for the model scale, so that the proposed federated recommendation approaches need to fine-tune the model on each device. 
Accordingly, limited device resources (e.g., storage, RAM, energy, and communication bandwidth, etc.) are heavily occupied. 
Moreover, existing federated approaches cannot effectively exploit collaborative filtering (CF) information among users/device, which limits the performance of existing federated recommendation methods.

To tackle the problems listed above, we focus on a new privacy-aware federated recommendation architecture for the \ac{RP} task~\citep{marlin2004modeling,koren2008factorization,li2017collaborative}. 
For the \ac{RP} task we aim to predict the rating that a user would give to an item that she has not rated in the past as precisely as possible~\citep{koren2009matrix,hu2014your}.
In this paper, our target is to design a novel federated learning framework to \ac{RP} for a federated mobile environment that operates on par with state-of-the-art fully centralized \ac{RP} methods. 

As the method of choice for the \ac{RP} task, \ac{MF} is used to optimize latent factors to represent users and items by projecting users and items into a joint dense vector space~\cite{koren2009matrix,he2017neural}.
Today's \ac{MF} methods consider \ac{RP} models as well as item embeddings of the same size and shared parameters for all users in order to predict personalized ratings.
For fitting all user data, the shared \ac{RP} model with item embeddings must be large in size.
In this paper, we hypothesize that using private item embeddings and models can achieve competitive performance with a small model scale, based on two intuitions.
First, different users might have different views and/or angles about the same item: it is not necessary for all users to use shared item embeddings that require many parameters.
Second, different users might favor different \ac{RP} strategies, which means we can use a specific and small model to fit a user's private data.
A key challenge is how we can build private \ac{RP} models on local devices and at the same time effectively utilize \ac{CF} information on the server as we may not have enough personal data for each user to build her own model.

In this paper, we address this challenge by introducing a novel matrix factorization framework, namely \acfi{MetaMF}.
Instead of building a model on each local device, we propose to ``generate'' private item embeddings and \ac{RP} models with a meta network.
Specifically, we assign a so-called indicator vector (i.e., a one-hot vector corresponding to a user id) to each user.
For a given user, we first fuse her indicator vector to get a collaborative vector by collecting useful information from other users with a \acf{CM} module.
Then, we employ a \acf{MR} module to generate private item embeddings and a \ac{RP} model based on the collaborative vector.
It is challenging to directly generate the item embeddings due to the large number of items and the high dimensions.
To tackle this problem, we devise a \acf{RG} strategy that first generates a low-dimensional item embedding matrix and a rise-dimensional matrix, and then multiply them to obtain high-dimensional embeddings.
Finally, we use the generated \ac{RP} model to obtain \acp{RP} for this user.
In a federated recommender system, we deploy the private \ac{RP} model on the user's device, and the meta network, including \ac{CM} and \ac{MR} modules, on the server.

We perform extensive experiments on four benchmark datasets.
Despite its federated nature, \ac{MetaMF} shows comparable performance with state-of-the-art \ac{MF} methods on two datasets, while using fewer parameters for item embeddings and \ac{RP} models.
Both the generated item embeddings and the \ac{RP} model parameters exhibit clustering phenomena, demonstrating that \ac{MetaMF} can effectively model \ac{CF} in a federated manner while generating a private model for each user. 
Moreover, we find that \ac{MetaMF} achieves higher \ac{RP} performance than state-of-the-art federated recommendation methods by better exploiting \ac{CF} among users/devices.
To facilitate reproducibility of the results, we are sharing the code.

The main contributions of this paper are as follows:
\begin{itemize}
    \item We introduce a novel federated \acf{MF} framework,  \ac{MetaMF}, that can reduce the parameters of \ac{RP} models and item embeddings without loss in performance.
    \item We devise a meta network, including \acl{CM} and \acl{MR} modules, to better exploit \acl{CF} in federated recommender systems.
    \item We propose a \acl{RG} strategy to reduce the parameters and calculation in generation.
    \item We conduct extensive experiments and analyses to verify the effectiveness and efficiency of \ac{MetaMF}.
\end{itemize}


\section{Related Work}

\begin{figure*}[t]
\centering
\includegraphics[width=1\linewidth]{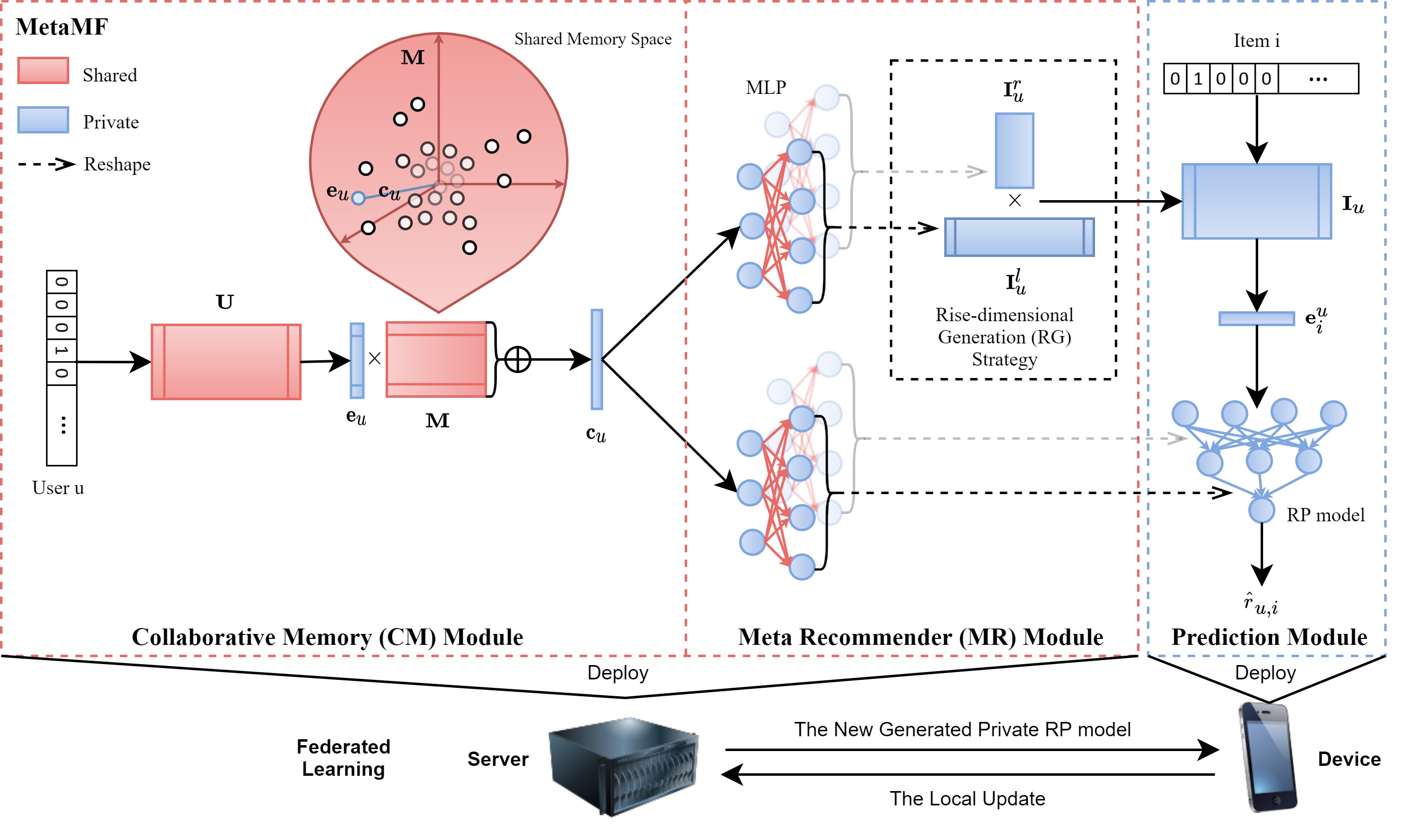}
\caption{An overview of \ac{MetaMF}. It consists of three modules. The \ac{CM} module and the \ac{MR} module with the \ac{RG} strategy tend to generate private item embeddings and \ac{RP} models for different users, which are deployed into the server. The prediction module aims to predict private ratings based on the generated item embeddings and \ac{RP} models for each user, which is deployed into the device.}
\label{fig_3_1}
\end{figure*}

We group related work into federated recommender systems, matrix factorization, and meta learning.

\subsection{Federated Recommender Systems}

For \acp{RP} in a federated environment, it is impractical to only rely on local data to train a model for each device, due to data sparsity.
Thus, previous work for the federated environment focuses on how to collaboratively train models on distributed data using existing recommendation methods.
\citet{ziegler2004semantic} propose to build a graph among computers based on trust, then to use \acl{CF} to do recommendations.
\citet{kermarrec2010application} further present a user-based random walk approach with \ac{CF} across devices to predict ratings.
\citet{DBLP:journals/corr/WangLCZQH15} introduce a parallel and distributed \acl{MF} algorithms to cooperatively learn user/item latent factors across multiple devices.
\citet{barbosa2018distributed} propose that smartphones exchange data between devices and calculate their own recommendation via collaborative filtering.
\citet{beierle_uic_2019} further present a mobile architecture consisting of data collection, data exchange, and a local recommender system; the data collection component gets data about the user from local device, data exchange gets data about other users from other devices, and the local recommender system utilizes all available data for recommending items to the user.

Several studies have introduced federated learning~\cite{mcmahan2017communication} into the realm of recommendation, which provides a way to realize federated recommender systems.
\citet{chen2018federated} propose a recommendation framework based on federated meta learning, which maintains a shared model in the cloud.
To adapt it for each user, they download the model to the local device and fine-tune the model for personalized recommendations.
\citet{ammad2019federated} formulate \ac{FCF} methods and adapt WRMF~\cite{hu2008collaborative} to demonstrate the applicability of \ac{FCF}.

Unlike us, \citet{ammad2019federated} do not focus on the size of the local models while maintaining performance; importantly, they focus on the ranking task, not the \acl{RP} task that we focus on.
In previous federated learning methods, the global model in the server and the local model in the device have the same size,  the local model is a copy of the global model.
No previous work uses the type of architecture that we design for \ac{MetaMF} that deploys a big meta network into the server to exploit \ac{CF}  while deploying a small \ac{RP} model into the device to predict ratings.

\subsection{Matrix Factorization}
Matrix factorization (\ac{MF}) has attracted a lot of attention since it was proposed for recommendation tasks.
Early studies focus mainly on how to achieve better rating matrix decomposition.
\citet{sarwar2000application} employ \ac{SVD} to reduce the dimensionality of the rating matrix, so that they can get low-dimensional user and item vectors.
\citet{goldberg2001eigentaste} apply \ac{PCA} to decompose the rating matrix, and obtain the principal components as user or item vectors.
\citet{zhang2006learning} propose \ac{NMF}, which decomposes the rating matrix by modeling each user's ratings as an additive mixture of rating profiles from user communities or interest groups and constraining the factorization to have non-negative entries.
\citet{mnih2008probabilistic} propose \ac{PMF} to model the distributions of user and item vectors from a probabilistic point of view.
\citet{koren2008factorization} proposes \ac{SVD}++, which enhances \ac{SVD} by including implicit feedback as opposed to \ac{SVD}, which only includes explicit feedback.

The matrix decomposition methods mentioned above estimate ratings by simply calculating the inner product between user and item vectors, which is not sufficient to capture their complex interactions.
Deep learning has been introduced to \ac{MF} to better model user-item interactions with non-linear transformations.
\citet{sedhain2015autorec} propose AutoRec, which takes ratings as input and reconstructs the ratings by an autoencoder.
Later, \citet{strub2016hybrid} enhance AutoRec by incorporating side information into a denoising autoencoder.
\citet{he2017neural} propose the \ac{NCF}, which employs a \ac{MLP} to model user-item interactions.
\citet{xue2017deep} present the \ac{DMF} which enhances \ac{NCF} by considering both explicit and implicit feedback.
\citet{He2018Outer} use \acp{CNN} to improve \ac{NCF} and present the ConvNCF, which uses the outer product to model user-item interactions.
\citet{cheng20183ncf} introduce an attention mechanism into \ac{NCF} to differentiate the importance of different user-item interactions.
Recently, a number of studies have investigated the use of side information or implicit feedback to enhance these neural models~\cite{li2017collaborative,xiao2018neural,xiao2019bayesian,yi2019deep}.

All these models provide personalized \acp{RP} by learning user representations to encode differences among users, while sharing item embeddings and models.
In contrast, \ac{MetaMF} provides private \acp{RP} by generating non-shared and small models as well as item embeddings for individual users.

\subsection{Meta Learning}
Meta learning, also known as ``learning to learn," has shown its effectiveness in reinforcement learning \cite{xu2018meta}, few-shot learning \cite{nichol2018first}, image classification \cite{ravi2016optimization}.

\citet{jia2016dynamic} propose a network to dynamically generate filters for \acp{CNN}.
\citet{bertinetto2016learning} introduce a model to predict the parameters of a pupil network from a single exemplar for one-shot learning.
\citet{ha2016hypernetworks} propose hypernetworks, which employ a network to generate the weights of another network.
\citet{krueger2017bayesian} present a Bayesian variant of hypernetworks that learns the distribution over the parameters of another network.
\citet{chen2018meta} use a hypernetwork to share function-level information across multiple tasks.
Few of them target recommendation, which is a more complex task with its own unique challenges.

Recently, some studies have introduced meta learning into recommendations.
\citet{vartak2017meta} study the item cold-start problem in recommendations from a meta learning perspective.
They view recommendation as a binary classification problem, where the class labels indicate whether the user engaged with the item.
Then they devise a classifier by adapting a few-shot learning paradigm \cite{snell2017prototypical}.
\citet{lee2019melu} propose a meta learning-based recommender
system called MeLU to alleviate the user cold-start problem.
MeLU can estimate new users' preferences with a few consumed items and determine distinguishing items for customized preference estimation by an evidence candidate selection strategy.
\citet{du2019sequential} unify scenario-specific learning and model-agnostic sequential meta learning into an integrated end-to-end framework, namely Scenario-specific Sequential Meta learner (s$^2$Meta).
s$^2$Meta can produce a generic initial model by aggregating contextual information from a variety of prediction tasks and effectively adapt to specific tasks by leveraging learning-to-learn knowledge.

Different from these publications, we learn a hypernetwork (i.e., \ac{MetaMF}) to directly generate private \ac{MF} models for each user for \acp{RP}.


\section{Meta Matrix Factorization}

\subsection{Overview}
Given a user $u$ and an item $i$, the goal of \acfi{RP} is to estimate a rating $\hat{r}_{u,i}$ that is as accurate as the true rating $r_{u,i}$.
We denote the set of users as $\mathcal{U}$, the set of items as $\mathcal{I}$, the set of true ratings as $\mathcal{R}$, which will be divided into the training set $D_{train}$, the validation set $D_{valid}$, and the test set $D_{test}$.

As shown in Fig.~\ref{fig_3_1}, \ac{MetaMF} has three components: a \textit{collaborative memory module} (see Section~\ref{sec:cmm}), a \textit{meta recommender module} (see Section~\ref{sec:mrm} and a \textit{prediction module} (i.e., a \ac{RP} model; see Section~\ref{sec:pm}), where the \ac{CM} and \ac{MR} modules constitute a meta network shared by all users, and the prediction module is private.
In \ac{CM} module, we first obtain the user embedding $\mathbf{e}_u$ of $u$ from the user embedding matrix $\mathbf{U}$ and take it as the coordinates to obtain the collaborative vector $\mathbf{c}_u$ from a shared memory space that fuses information from all users.
Then we input $\mathbf{c}_u$ to the \ac{MR} module to generate the parameters of a private \ac{RP} model for $u$.
The \ac{RP} model can be of any type.
In this work, the \ac{RP} model is a \acf{MLP}.
We also generate the private item embedding matrix $\mathbf{I}_u$ of $u$ with a \textit{rise-dimensional generation strategy}.
Finally, the prediction module takes the item embedding $\mathbf{e}_{ui}$ of $i$ from $\mathbf{I}_u$ as input and predicts $r_{u,i}$ using the generated \ac{RP} model.

\subsection{Federated Rating Predictions}

Before we detail each module, we first detail how to use \ac{MetaMF} to decentralize data to build a federated recommender system.
Because \ac{MetaMF} can be divided into a meta network, including \ac{CM} and \ac{MR} modules, and a \ac{RP} model, i.e., the prediction module, making it suitable to combine with federated learning to realize a federated recommender system.

Specifically, we can deploy the \ac{CM} and \ac{MR} modules into a data center, i.e., the server, and deploy the prediction module locally into mobile devices.
The centralized server first generates and delivers different parameters to different mobile devices.
Next, each mobile device calculates the loss and the gradients of the parameters in the prediction module based on its private data, and uploads the gradients to the server.
Then the server can calculate the gradients of the parameters in the \ac{CM} and \ac{MR} modules based on the gradients gathered from each device, and update the parameters.
Finally, the server generates and delivers new parameters to each mobile device.
Like federated machine learning methods, \ac{MetaMF} can protect user privacy to a certain extent, because user data does not need to be uploaded to the server.
Naturally, the strength of the privacy protection depends on the content of the updates; see Section~\ref{section:conclusion}.

\ac{MetaMF} provides a solid trade-off between exploiting \ac{CF} for higher RP performance and protecting users' personal information.
It places the meta network with the most parameters in the server and places the prediction module of a small scale in devices, which is more suitable to a mobile environment with limited storage, RAM, energy and communication bandwidth.

\subsection{Collaborative Memory Module}
\label{sec:cmm}

In order to facilitate collaborative filtering, we propose the \ac{CM} module to learn a collaborative vector for each user, which encodes both the user's own information and some useful information from other users.

Specifically, we assign each user $u$ and each item $i$ the indicator vectors, $\mathbf{i}_u\in\mathbb{R}^m$ and $\mathbf{i}_i\in\mathbb{R}^n$ respectively, where $m$ is the number of users and $n$ is the number of items.
Note that $\mathbf{i}_u$ and $\mathbf{i}_i$ are one-hot vectors with each dimension corresponding to a particular user or item.
For the given user $u$, we first get the user embedding $\mathbf{e}_u$ by Eq.~\ref{e_u}:
\begin{equation}
\label{e_u}
\mathbf{e}_u = \mathbf{U}\mathbf{i}_u,
\end{equation}
where $\mathbf{e}_u\in\mathbb{R}^{d_u}$, $\mathbf{U}\in\mathbb{R}^{d_u\times{m}}$ is the user embedding matrix, and $d_u$ is the size of user embeddings.
Then we proceed to obtain a collaborative vector for $u$.
Specifically, we use a shared memory matrix $\mathbf{M}\in\mathbb{R}^{d_u\times{k}}$ to store the basis vectors which span a space of all collaborative vectors, where $k$ is the dimension of basis vectors and collaborative vectors.
And we consider the user embedding $\mathbf{e}_u$ as the coordinates of $u$ in the shared memory space.
So the collaborative vector $\mathbf{c}_u\in\mathbb{R}^k$ for $u$ is a linear combination of the basis vectors in $\mathbf{M}$ by $\mathbf{e}_u$, as shown in Eq.~\ref{c_u}:
\begin{equation}
\label{c_u}
\mathbf{c}_u = \sum_i\mathbf{M}(i,:)\mathbf{e}_u(i),
\end{equation}
where $\mathbf{M}(i,:)$ is the $i$-th vector of $\mathbf{M}$ and $\mathbf{e}_u(i)$ is the $i$-th scalar of $\mathbf{e}_u$.
Because the memory matrix $\mathbf{M}$ is shared among all users, the shared memory space will fuse information from all users.
\ac{MetaMF} can flexibly exploit collaborative filtering among users by assigning them with similar collaborative vectors in the space defined by $\mathbf{M}$, which is equivalent to learning similar user embeddings as in existing \ac{MF} methods.

\subsection{Meta Recommender Module}
\label{sec:mrm}

We propose the \ac{MR} module to generate the private item embeddings and \ac{RP} model based on the collaborative vector from the \ac{CM} module.

\subsubsection{Private Item Embeddings.}
We propose to generate the private item embedding matrix $\mathbf{I}_u\in\mathbb{R}^{d_i\times{n}}$ for each user $u$, where $d_i$ is the size of item embeddings.
However, it is a challenge to directly generate the whole item embedding matrix when there are a large number of items with relatively high-dimensional item embeddings (instead of extremely small ones).
Therefore, we propose a \acf{RG} strategy to decompose the generation into two parts: a low-dimensional item embedding matrix $\mathbf{I}^l_u\in\mathbb{R}^{s\times{n}}$ and a rise-dimensional matrix $\mathbf{I}^r_u\in\mathbb{R}^{d_i\times{s}}$, where $s$ is the size of low-dimensional item embeddings and $s \ll d_i$.
Specifically, we first follow Eq.~\ref{I_l_r} to generate $\mathbf{I}^l_u\in\mathbb{R}^{sn}$ and $\mathbf{I}^r_u\in\mathbb{R}^{d_is}$ (in the form of vectors):
\begin{equation}
\begin{split}
\label{I_l_r}
& \mathbf{h}^l_i = \mathrm{ReLU}(\mathbf{W}^l_i\mathbf{c}_u+\mathbf{b}^l_i), \quad \mathbf{I}^l_u = \mathbf{U}^l_i\mathbf{h}^l_i, \\
& \mathbf{h}^r_i = \mathrm{ReLU}(\mathbf{W}^r_i\mathbf{c}_u+\mathbf{b}^r_i), \quad \mathbf{I}^r_u = \mathbf{U}^r_i\mathbf{h}^r_i,
\end{split}
\end{equation}
where $\mathbf{W}^l_i$ and $\mathbf{W}^r_i\in\mathbb{R}^{o\times{k}}$, $\mathbf{U}^l_i\in\mathbb{R}^{sn\times{o}}$ and  $\mathbf{U}^r_i\in\mathbb{R}^{d_is\times{o}}$ are weights; $\mathbf{b}^l_i$ and $\mathbf{b}^r_i\in\mathbb{R}^o$ are biases; $\mathbf{h}^l_i$ and $\mathbf{h}^r_i\in\mathbb{R}^o$ are hidden states;
$o$ is the hidden size.
Then we reshape $\mathbf{I}^l_u$ to a matrix whose shape is $s\times{n}$, and reshape $\mathbf{I}^r_u$ to a matrix whose shape is $d_i\times{s}$.
Finally, we multiply $\mathbf{I}^l_u$ and $\mathbf{I}^r_u$ to get $\mathbf{I}_u$:
\begin{equation}
\mathbf{I}_u = \mathbf{I}^r_u\mathbf{I}^l_u.
\end{equation}
Compared to directly generating $\mathbf{I}_u$, which needs $O(d_i\times{n})$ parameters, the \ac{RG} strategy needs $O(s\times{n}+d_i\times{s})$ parameters which reduces the cost of generating $\mathbf{I}_u$.
For different users, the generated item embedding matrices are different.

\subsubsection{Private \ac{RP} Model.}
We also propose to generate a private \ac{RP} model for each user $u$.
We use a \ac{MLP} as the \ac{RP} model, so we need to generate the weights and biases for each layer of \ac{MLP}.
Specifically, for layer $l$, we denote its weights and biases as $\mathbf{W}^u_l\in\mathbb{R}^{f_\mathit{out}\times{f_\mathit{in}}}$ and $\mathbf{b}^u_l\in\mathbb{R}^{f_\mathit{out}}$ respectively, where $f_\mathit{in}$ is the size of its input and $f_\mathit{out}$ is the size of its output.
Then $\mathbf{W}^u_l$ and $\mathbf{b}^u_l$ are calculated as follows:
\begin{equation}
\begin{split}
\label{l_W_b}
\mathbf{h}_g = {} & \mathrm{ReLU}(\mathbf{W}^h_g\mathbf{c}_u+\mathbf{b}^h_g), \\
\mathbf{W}^u_l = {} & \mathbf{U}^w_g\mathbf{h}_g+\mathbf{b}^w_g, \\
\mathbf{b}^u_l = {} & \mathbf{U}^b_g\mathbf{h}_g+\mathbf{b}^b_g,
\end{split}
\end{equation}
where $\mathbf{W}^h_g\in\mathbb{R}^{o\times{k}}$, $\mathbf{U}^w_g\in\mathbb{R}^{f_\mathit{out}f_{in}\times{o}}$ and $\mathbf{U}^b_g\in\mathbb{R}^{f_\mathit{out}\times{o}}$ are weights;
$\mathbf{b}^h_g\in\mathbb{R}^o$, $\mathbf{b}^w_g\in\mathbb{R}^{f_\mathit{out}f_\mathit{in}}$ and $\mathbf{b}^b_g\in\mathbb{R}^{f_\mathit{out}}$ are biases; $\mathbf{h}_g\in\mathbb{R}^o$ is hidden state.
Finally, we reshape $\mathbf{W}^u_l$ to a matrix whose shape is $f_\mathit{out}\times{f_\mathit{in}}$.
Note that $\mathbf{W}^h_g$, $\mathbf{b}^h_g$, $\mathbf{U}^w_g$, $\mathbf{b}^w_g$, $\mathbf{U}^b_g$ and $\mathbf{b}^b_g$ are not shared by different layers of the \ac{RP} model.
And $f_\mathit{in}$ and $f_\mathit{out}$ also vary with different layers.
Detailed settings can be found in the experimental setup.
Also, \ac{MetaMF} returns different parameters of the \ac{MLP} to each user.

\subsection{Prediction Module}
\label{sec:pm}

The prediction module estimates the user's rating for a given item $i$ using the generated item embedding matrix $\mathbf{I}_u$ and \ac{RP} model from the \ac{MR} module.

First, we get the private item embedding $\mathbf{e}^u_i\in\mathbb{R}^{d_i}$ of $i$ from $\mathbf{I}_u$ by Eq.~\ref{e_i}:
\begin{equation}
\label{e_i}
\mathbf{e}^u_i = \mathbf{I}_u\mathbf{i}_i.
\end{equation}
Then we follow Eq.~\ref{r_ui} to predict $r_{u,i}$ based on the \ac{RP} model:
\begin{equation}
\begin{split}
\label{r_ui}
\mathbf{h}_1 = {} & \mathrm{ReLU}(\mathbf{W}^u_1\mathbf{e}^u_i+\mathbf{b}^u_1), \\
\mathbf{h}_2 = {} & \mathrm{ReLU}(\mathbf{W}^u_2\mathbf{h}_1+\mathbf{b}^u_2), \\
\vdots\makebox[1.95mm]{} & \\
\mathbf{h}_{L-1} = {} & \mathrm{ReLU}(\mathbf{W}^u_{L-1}\mathbf{h}_{L-2}+\mathbf{b}^u_{L-1}), \\
\hat{r}_{u,i} = {} & \mathbf{W}^u_L\mathbf{h}_{L-1}+\mathbf{b}^u_L,
\end{split}
\end{equation}
where $L$ is the number of layers of the \ac{RP} model.
The weights $\{\mathbf{W}^u_1, \mathbf{W}^u_2, \ldots, \mathbf{W}^u_{L-1}, \mathbf{W}^u_L\}$ and biases $\{\mathbf{b}^u_1, \mathbf{b}^u_2, \ldots, \mathbf{b}^u_{L-1}, \mathbf{b}^u_L\}$ are generated by the \ac{MR} module.
The last layer $L$ is the output layer, which returns a scalar as the predicted rating $\hat{r}_{u,i}$.

\subsection{Loss}
In order to learn \ac{MetaMF}, we formulate the \ac{RP} task as a regression problem and the loss function is defined as:
\begin{equation}
\label{L_rp}
L_\mathit{rp} = \frac{1}{|D_\mathit{train}|} \sum_{r_{u,i}\in{D_\mathit{train}}}(r_{u,i}-\hat{r}_{u,i})^2.
\end{equation}
To avoid overfitting, we add the L2 regularization term:
\begin{equation}
L_\mathit{reg} = \frac{1}{2}\|\Theta\|^2_2,
\end{equation}
where $\Theta$ represents the trainable parameters of \ac{MetaMF}.
Note that unlike existing \ac{MF} methods, the item embeddings and the parameters of \ac{RP} models are not included in $\Theta$, because they are also the outputs of \ac{MetaMF}, not trainable parameters.

The final loss $L$ is a linear combination of $L_\mathit{rp}$ and $L_\mathit{reg}$:
\begin{equation}
L = L_\mathit{rp}+\lambda L_\mathit{reg},
\end{equation}
where $\lambda$ is the weight of $L_\mathit{reg}$.
The whole framework of \ac{MetaMF} can be efficiently trained using back propagation with federated learning on decentralized data, as showed in Algorithm~\ref{alg_1}.

\begin{algorithm}[t]
\caption{MetaMF}
\begin{algorithmic}[1]
\Require 
All trainable parameters $\Theta$, which are stored in the server;
The user set $\mathcal{U}$, where one user per device;
For user $u$, her local data $D_u$ stored in her device, and $D_\mathit{train}=\sum_{u\in\mathcal{U}}D_u$; $^\dagger$ means the code is executed in the device.
\Ensure
$\Theta$;
For user $u$, the parameters of her \ac{RP} model and item embeddings $\Phi_u$, which are stored in her device;
\State Initialize $\Theta$ randomly in the server;
\For{u in $\mathcal{U}$}
    \State Generate $\Phi_u$ based on $\Theta$;
    \State Send $\Phi_u$ to $u$'s device;
\EndFor
\While{not convergent}
    \State Sample a batch $S$ from $\mathcal{U}$;
    \For{$u$ in $S$}
        \State Sample a batch $B_u$ from $D_u$;$^\dagger$ 
        \State Calculate the gradient of $\Phi_u$ based on $B_u$;$^\dagger$ 
        \State Upload the gradient to the server;$^\dagger$
        \State Calculate the gradient of $\Theta$ based on the gradient of $\Phi_u$;
    \EndFor
    \State Accumulate the gradients of $\Theta$ gathered from $S$;
    \State Update $\Theta$ based on the accumulated gradient;
    \For{u in $\mathcal{U}$}
        \State Regenerate $\Phi_u$ based on new $\Theta$;
        \State Send $\Phi_u$ to $u$'s device;
    \EndFor    
\EndWhile
\end{algorithmic}
\label{alg_1}
\end{algorithm}

\section{Experimental Setup}
\label{section:experimental setup}

\begin{table}[htb]
\caption{Statistics of the datasets, where \textit{\#avg} means the average number of user ratings, Hetrec-ML is the short name of Hetrec-movielens.}
\label{table_4_1}
\centering
\setlength\tabcolsep{3pt}
\begin{tabular}{l rrrrr}
\toprule
\textbf{Datasets} &  \bf\textit{\#users} & \bf\textit{\#items} & \bf\textit{\#ratings} & \bf\textit{\#avg} & \bf\textit{\#sparsity (\%)}\\
\midrule
Douban & 2,509 & 39,576 & 894,887 & 357 & 0.9\phantom{0} \\
Hetrec-ML & 2,113 & 10,109 & 855,599 & 405 & 4\phantom{.00} \\
Movielens1M & 6,040 & 3,706 & 1,000,209 & 166 & 4.5\phantom{0} \\
Ciao & 7,375 & 105,096 & 282,619 & 38 & 0.04 \\
\bottomrule
\end{tabular}
\squeeze
\end{table}

\begin{table*}
\caption{Comparison results of \ac{MetaMF} and baselines on the four datasets. A superscript $^{\approx}$ indicates that there is no statistically significant difference between \ac{MetaMF} and NCF (two-sided paired t-test, $p<0.01$).}
\label{table_5_1}
\centering
\setlength\tabcolsep{12pt}
\begin{tabular}{l cc cc cc cc}
\toprule
\multirow{2}{*}{\bf Method} & \multicolumn{2}{c}{\bf Douban} & \multicolumn{2}{c}{\bf Hetrec-movielens} & \multicolumn{2}{c}{\bf Movielens1M} & \multicolumn{2}{c}{\bf Ciao}\\
\cmidrule(r){2-3} \cmidrule(r){4-5} \cmidrule(r){6-7} \cmidrule(r){8-9}
& \multicolumn{1}{c}{\bf MAE} & \multicolumn{1}{c}{\bf MSE} & \multicolumn{1}{c}{\bf MAE} & \multicolumn{1}{c}{\bf MSE} & \multicolumn{1}{c}{\bf MAE} & \multicolumn{1}{c}{\bf MSE} & \multicolumn{1}{c}{\bf MAE} & \multicolumn{1}{c}{\bf MSE} \\
\midrule
NMF & 0.602 & 0.585 & 0.625 & 0.676 & 0.727 & 0.848 & 0.750 & 1.039 \\
PMF & 0.639 & 0.701 & 0.617 & 0.644 & 0.703 & 0.788 & 1.501 & 3.970 \\
SVD++ & 0.593 & 0.570 & 0.579 & 0.590 & 0.671 & 0.740 & 0.738 & 0.963 \\
LLORMA & 0.610 & 0.623 & 0.588 & 0.603 & 0.675 & 0.748 & 1.349 & 3.396 \\
\midrule
RBM & 1.058 & 1.749 & 1.124 & 1.947 & 1.122 & 2.078 & 1.132 & 2.091 \\
AutoRec-U & 0.709 & 0.911 & 0.660 & 0.745 & 0.678 & 0.775 & 1.673 & 5.671  \\
AutoRec-I & 0.704 & 0.804 & 0.633 & 0.694 & \textbf{0.663} & \textbf{0.715} & 0.792 & 1.038 \\
NCF & \textbf{0.583} & \textbf{0.547} & 0.572 & \textbf{0.575} & 0.675 & 0.739 & \textbf{0.735} & \textbf{0.937} \\
\midrule
FedRec & 0.760 & 0.927 & 0.846 & 1.265 & 0.907 & 1.258 & 0.865 & 1.507 \\
\midrule
MetaMF & 0.584\rlap{$^{\approx}$} & 0.549 & \textbf{0.571}\rlap{$^{\approx}$} &  0.578\rlap{$^{\approx}$} & 0.687 & 0.760 & 0.774 & 1.043 \\
\bottomrule
\end{tabular}
\end{table*}

\begin{table*}[!t]
\caption{Rating prediction results of \ac{MetaMF}, \ac{MetaMF}-SI and \ac{MetaMF}-SM on the four datasets. \ac{MetaMF}-SI shares item embeddings for all users; \ac{MetaMF}-SM shares the parameters of prediction module for all users.}
\label{table_5_2}
\centering
\setlength\tabcolsep{12pt}
\begin{tabular}{l cc cc cc cc}
\toprule
\multirow{2}{*}{\bf Method} & \multicolumn{2}{c}{\bf Douban} & \multicolumn{2}{c}{\bf Hetrec-movielens} & \multicolumn{2}{c}{\bf Movielens1M} & \multicolumn{2}{c}{\bf Ciao} \\
\cmidrule(r){2-3} \cmidrule(r){4-5} \cmidrule(r){6-7} \cmidrule{8-9} 
& \multicolumn{1}{c}{\bf MAE} & \multicolumn{1}{c}{\bf MSE} & \multicolumn{1}{c}{\bf MAE} & \multicolumn{1}{c}{\bf MSE} & \multicolumn{1}{c}{\bf MAE} & \multicolumn{1}{c}{\bf MSE} & \multicolumn{1}{c}{\bf MAE} & \multicolumn{1}{c}{\bf MSE} \\
\midrule
MetaMF & \textbf{0.584} & \textbf{0.549} & \textbf{0.571} & \textbf{0.578} & \textbf{0.687} & \textbf{0.760} & 0.774 & 1.043 
\\
MetaMF-SI & 0.586 & 0.552 & 0.590 & 0.615 & 0.696 & 0.784 & \textbf{0.732} & \textbf{0.925} \\
MetaMF-SM & 0.595 & 0.571 & 0.595 & 0.622 & 0.697 & 0.788 & 0.789 & 1.061 
\\
\bottomrule
\end{tabular}
\end{table*}

We seek to answer the following research questions.
(RQ1)~How does the proposed method \ac{MetaMF} for federated \aclp{RP} perform compared to state-of-the-art \ac{MF} methods for the \ac{RP} task? Does the federated nature of \ac{MetaMF} come at cost in terms of performance on the \ac{RP} task?
(RQ2)~What is the contribution of generating private item embeddings and \ac{RP} models?

\subsection{Datasets}
We conduct experiments on four widely used datasets: \textbf{Douban}~\cite{hu2014your}, \textbf{Hetrec-movielens}~\cite{cantador2011second}, \textbf{Movielens1M}~\citep{harper2016movielens} and \textbf{Ciao}~\citep{guo2014etaf}.
We list the statistics of these four datasets in Table~\ref{table_4_1}.
For each user on each dataset, we randomly separate her data into three chunks: $80\%$ as the training set, $10\%$ as the validation set and $10\%$ as the test set.

\subsection{Baselines}
We compare \ac{MetaMF} with the following conventional, deep learning-based and federated \ac{MF} methods.
It is worth noting that in this paper we focus on predicting ratings based on rating matrices, thus for fairness we neglect \ac{MF} methods that need side information.

\begin{itemize}
\item \textbf{Conventional methods:}
    \begin{itemize}
    \item \textbf{NMF}~\cite{zhang2006learning}: uses non-negative matrix factorization to decompose rating matrices.
    \item \textbf{PMF}~\cite{mnih2008probabilistic}: applies Gaussian distributions to model the latent factors of users and items.
    \item \textbf{SVD++}~\cite{koren2008factorization}: extends \ac{SVD} by considering implicit feedback for modeling latent factors.
    \item \textbf{LLORMA}~\cite{lee2016llorma}: uses a number of low-rank sub-matrices to compose rating matrices.
    \end{itemize}
\item \textbf{Deep learning-based methods:}
    \begin{itemize}
    \item \textbf{RBM}~\cite{salakhutdinov2007restricted}: employs \ac{RBM} to model the generation process of ratings.
    \item \textbf{AutoRec}~\cite{sedhain2015autorec}: proposes \acp{AE} to model interactions between users and items. AutoRec has two variants, with one taking users' ratings as input, denoted by AutoRec-U, and the other taking items' ratings as input, denoted by AutoRec-I.
    \item \textbf{NCF}~\cite{he2017neural}: the state-of-the-art \ac{MF} method that combines generalized matrix factorization and \ac{MLP} to model user-item interactions. We adapt \ac{NCF} for the \ac{RP} task by dropping the sigmoid activation function on its output layer and replacing its loss function with Eq.~\ref{L_rp}.
    \end{itemize}
\item \textbf{Federated methods:}
    \begin{itemize}
    \item \textbf{FedRec}~\cite{chen2018federated}: a federated recommendation method, which employs MAML~\cite{finn2017model} to learn a shared \ac{RP} model in the server and update the model for each device.
    In our experiments, the shared \ac{RP} model is a \ac{MLP} with two layers (layer sizes are $16$ and $1$ respectively), and its user/item embedding size is $64$.
    \end{itemize}
\end{itemize}

\subsection{Evaluation Metrics}
To evaluate the performance of rating prediction methods, we employ two evaluation metrics, i.e., \ac{MAE} and \ac{MSE}.
Both of them are widely applied for the \ac{RP} task in recommender systems.
Given the predicted rating $\hat{r}_{u,i}$ and the true rating $r_{u,i}$ of user $u$ on item $i$ in the test set $D_\mathit{test}$, \ac{MAE} is calculated as:
\begin{equation}
\ac{MAE} = \frac{1}{|D_\mathit{test}|}\sum_{r_{u,i}\in{D_\mathit{test}}}|r_{u,i}-\hat{r}_{u,i}|.
\end{equation}
\ac{MSE} is defined as:
\begin{equation}
\ac{MSE} = \frac{1}{|D_\mathit{test}|}\sum_{r_{u,i}\in{D_\mathit{test}}}(r_{u,i}-\hat{r}_{u,i})^2.
\end{equation}
Statistical significance of observed differences is tested for using a two-sided paired t-test for significant differences ($p<0.01$).

\subsection{Implementation Details}
The user embedding size $d_u$ and the item embedding size $d_i$ are set to $32$.
The size of the collaborative vector $k$ is set to $128$.
The size of the low-dimensional item embedding $s$ is set to $8$.
The hidden size $o$ is set to $512$.
And the \ac{RP} model in the prediction module is an \ac{MLP} with two layers (one hidden layer and one output layer) whose layer sizes are $8$ and $1$.
During training, we initialize all trainable parameters randomly with the Xavier method~\citep{Glorot2010Understanding}.
We choose Adam~\citep{Kingma2014Adam} to optimize \ac{MetaMF}, set the learning rate to $0.0001$, and set the regularizer weight $\lambda$ to $0.001$.
Our framework is implemented with Pytorch~\cite{paszke2019pytorch}.
In our experiments, we implement \ac{NCF} based on the released code of the author.\footnote{\url{https://github.com/hexiangnan/neural_collaborative_filtering}}
We use the code released by the respective authors\footnote{\url{https://github.com/gtshs2/Autorec}} for AutoRec. 
We use Lib\-Rec\footnote{\url{https://www.librec.net/}} for the remaining baselines.

\section{Experimental Results}
\label{section:experimentalresults}

\subsection{What Is the Cost of Federation?}
\label{RQ1}

We start by addressing RQ1 and compare our federated \acl{RP} model \ac{MetaMF} with state-of-the-art \ac{MF} methods.
Table~\ref{table_5_1} lists the \ac{RP} performance of all MF methods.

Our main observations are as follows.
First, on the Douban and Hetrec-movielens datasets, \ac{MetaMF} outperforms most baselines despite the fact that it is federated while most baselines are centralized.
And \ac{MetaMF} is slightly inferior to \ac{NCF}, but this difference is not significant.
So we can draw the conclusion that the performance of \ac{MetaMF} is comparable to \ac{NCF} on these two datasets.
See Section~\ref{RQ2} and Section~\ref{RQ3} for further analysis.

Second, on the Movielens1M and Ciao datasets, \ac{MetaMF} does not perform well, in some cases worse than some traditional methods, such as \ac{SVD}++.
The most important reason is that the average numbers of user ratings on these two datasets are small.
As shown in Table~\ref{table_4_1}, the statistics \textit{\#avg} on these four datasets are $357$, $405$, $166$ and $38$ respectively.
Because the Douban and Hetrec-movielens datasets provide more private data for each user, \ac{MetaMF} is able to capture the differences among users for learning private item embeddings and \ac{RP} models.
However, the Movielens1M and Ciao datasets lack sufficient data, which limits the performance of \ac{MetaMF}.

Third and finally, \ac{MetaMF} significantly outperforms FedRec on all datasets with smaller user/item embedding size and \ac{RP} model scale.
And FedRec performs worse than most baselines on most datasets.
The reason may be that FedRec cannot effectively exploit \ac{CF} information among users/devices.
Although FedRec maintains a shared model in the server, it needs to fine-tune the model on each device, which prevents some useful information from being shared among devices.
However, \ac{MetaMF} can flexibly take advantage of \ac{CF} among users/devices by the meta network.
See Section~\ref{RQ4} for further analysis.

Although federated recommender systems can protect user privacy by keeping data locally, it is harder for them to exploit \acl{CF} among users than for centralized approaches, which affects their performance.
So how to share information among multiple devices in a privacy-aware manner is still a core problem in federated recommender systems.
We can observe that \ac{MetaMF} does not outperform \ac{NCF}, and the performance of FedRec is also worse than of most centralized baselines.
Although the federated nature of \ac{MetaMF} makes it trade performance for privacy, it can still achieve comparable performance with \ac{NCF} on two datasets, which shows \ac{MetaMF} can get a better balance between privacy protection and \ac{RP} performance.

\subsection{What Does the Privatization of \ac{MetaMF} Contribute?}
\label{RQ2}

Next we address RQ2 to analyze the effectiveness of generating private item embeddings and \ac{RP} models to the overall performance of \ac{MetaMF}.
First, we compare \ac{MetaMF} to \ac{MetaMF}-SI, which only generates private \ac{RP} models for different users while sharing a common item embedding matrix among all users.
As shown in Table~\ref{table_5_2}, \ac{MetaMF} outperforms \ac{MetaMF}-SI on most datasets, except for the Ciao dataset.
We conclude that generating private item embeddings for each user can improve the performance of \ac{MetaMF}.
It is possible that the Ciao dataset lacks sufficient private data for learning private item embeddings, so that \ac{MetaMF} performs worse than \ac{MetaMF}-SI.
And if we compare \ac{MetaMF}-SI with \ac{NCF}, we find that \ac{MetaMF}-SI also outperforms \ac{NCF} on the Ciao dataset, which indicates that generating private \ac{RP} models can improve \ac{RP} on the Ciao dataset.

Next, we compare \ac{MetaMF} with \ac{MetaMF}-SM, which generates different item embeddings for different users and shares a common \ac{RP} model among all users.
From Table~\ref{table_5_2}, we can see that \ac{MetaMF} consistently outperforms \ac{MetaMF}-SM on all datasets.
Thus, generating private \ac{RP} models for users is able to improve the performance of \ac{MetaMF} too.

Furthermore, by comparing \ac{MetaMF}-SI and \ac{MetaMF}-SM, we observe that \ac{MetaMF}-SI outperforms \ac{MetaMF}-SM on all datasets.
This shows that item embeddings have a greater impact on the performance of \ac{MetaMF} than \ac{RP} models.

Finally, in response to RQ2 we conclude that generating private item embeddings and \ac{RP} models in \ac{MetaMF} contributes to overall performance of \ac{MetaMF}.


\section{Analysis}
\label{section:analysis}

\begin{table}[t]
\caption{The performance of \ac{MetaMF} with different model scales, where each model scale is represented as a tuple (\emph{item} \emph{embedding size}, [\emph{layer sizes in the Prediction Module}], [\emph{collaborative vector size}, \emph{hidden layer size in the \ac{MR} Module}]).}
\label{table_5_3}
\centering
\begin{tabular}{l @{~} cc cc}
\toprule
\multirow{2}{*}{\bf Model scale} & \multicolumn{2}{c}{\bf Douban} & \multicolumn{2}{c}{\bf Hetrec-movielens} \\
\cmidrule(r){2-3} \cmidrule{4-5} 
& \multicolumn{1}{c}{\bf MAE} & \multicolumn{1}{c}{\bf MSE} & \multicolumn{1}{c}{\bf MAE} & \multicolumn{1}{c}{\bf MSE} \\
\midrule
$(8,[2,1],[32,128])$ & 0.584 & \textbf{0.548} & 0.575 & 0.584 \\
$(16,[4,1],[64,256])$ & 0.587 & 0.552 & 0.573 & 0.582 \\
$(32,[8,1],[128,512])$ & \textbf{0.584} & 0.549 & \textbf{0.571} & \textbf{0.578} \\
$(64,[16,1],[256,1024])$ & --- & --- & 0.578 & 0.591 \\
\bottomrule
\end{tabular}
\bigskip
\caption{The performance of \ac{NCF} with different model scales, where each model scale is represented as (\emph{item embedding size}, [\emph{layer sizes}]).}
\label{table_5_4}
\centering
\begin{tabular}{l cc cc}
\toprule
\multirow{2}{*}{\bf Model scale} & \multicolumn{2}{c}{\bf Douban} & \multicolumn{2}{c}{\bf Hetrec-movielens} \\
\cmidrule(r){2-3} \cmidrule{4-5} 
& \multicolumn{1}{c}{\bf MAE} & \multicolumn{1}{c}{\bf MSE} & \multicolumn{1}{c}{\bf MAE} & \multicolumn{1}{c}{\bf MSE} \\
\midrule
$(16,[16,8,4,1])$ & 0.587 & 0.552 & 0.585 & 0.603 \\
$(32,[32,16,8,1])$ & 0.587 & 0.552 & 0.583 & 0.600 \\
$(64,[64,32,16,1])$ & 0.584 & 0.549 & 0.579 & 0.595 \\
$(128,[128,64,32,1])$ & 0.585 & 0.549 & 0.574 & 0.579 \\
$(256,[256,128,64,1])$ & \textbf{0.583} & \textbf{0.547} & \textbf{0.572} & \textbf{0.575} \\
$(512,[512,256,128,1])$ & 0.584 & 0.547 & 0.572 & 0.581 \\
$(1024,[1024,512,256,1])$ & 0.586 & 0.549 & 0.574 & 0.582 \\
\bottomrule
\end{tabular}
\end{table}

\begin{figure*}[t]
\centering
\subfigure[The weights of the hidden layer on the Douban dataset.]{
\includegraphics[width=0.2381\linewidth]{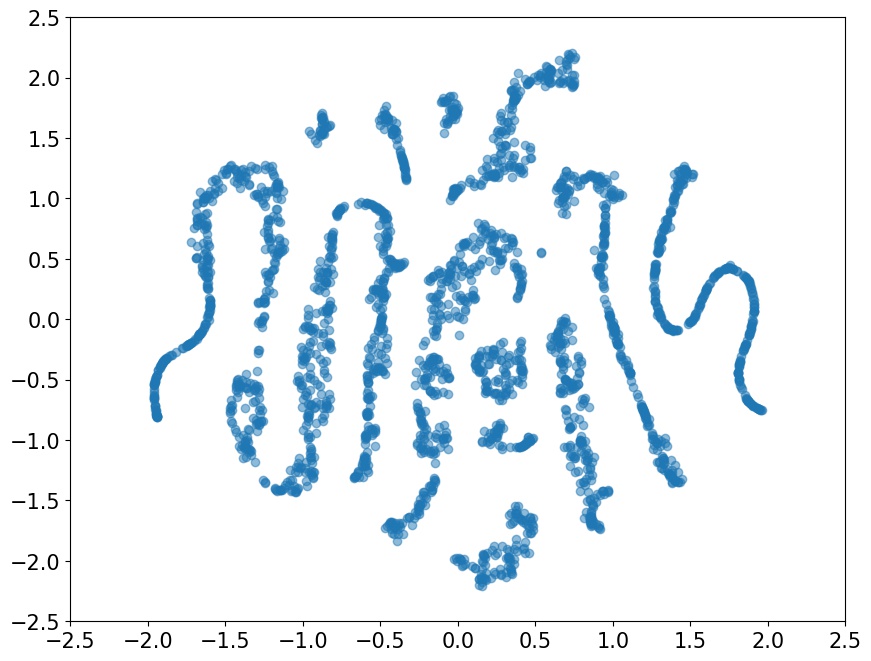}
}
\subfigure[The embeddings of item 16716 on the Douban dataset.]{
\includegraphics[width=0.2381\linewidth]{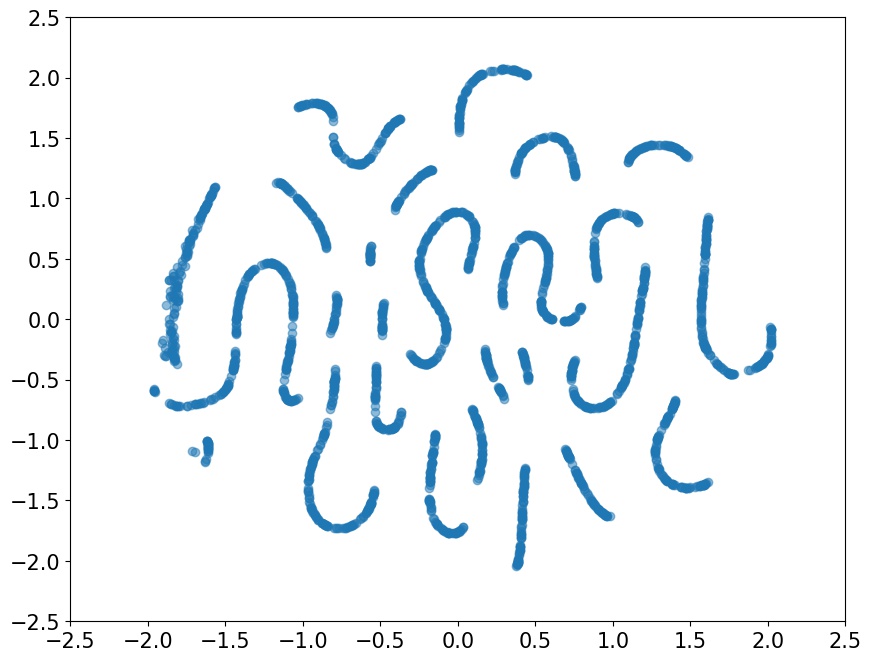}
}
\subfigure[The weights of the hidden layer on the Hetrec-movielens dataset.]{
\includegraphics[width=0.2381\linewidth]{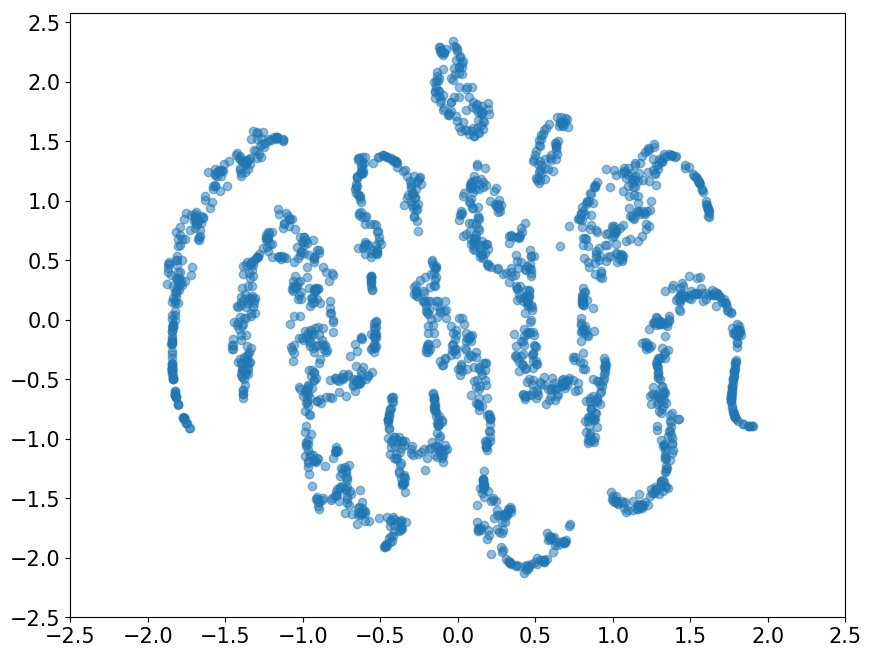}
}
\subfigure[The embeddings of item 1931 on the Hetrec-movielens dataset.]{
\includegraphics[width=0.2381\linewidth]{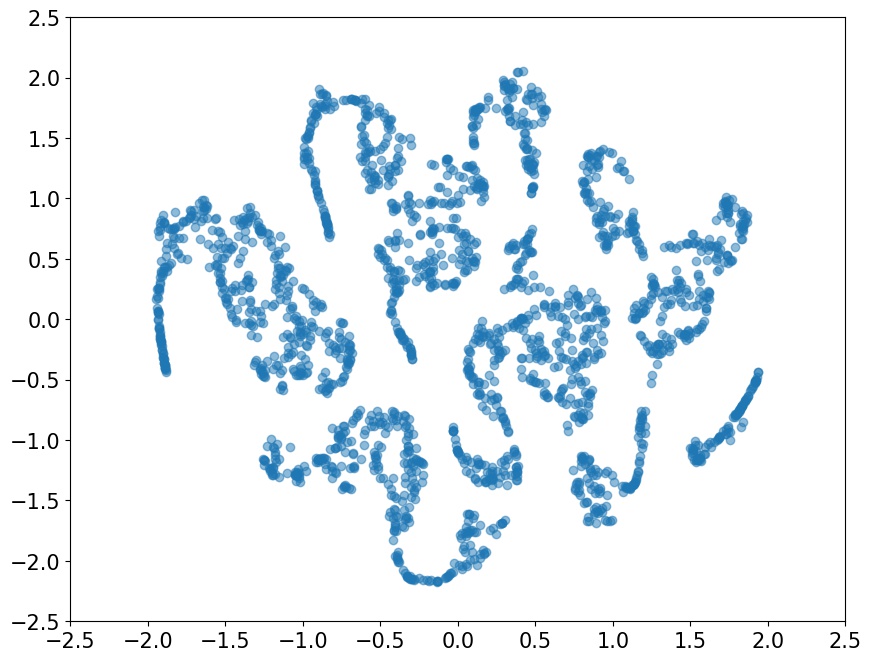}
}
\subfigure[The weights of the hidden layer on the Movielens1M dataset.]{
\includegraphics[width=0.2381\linewidth]{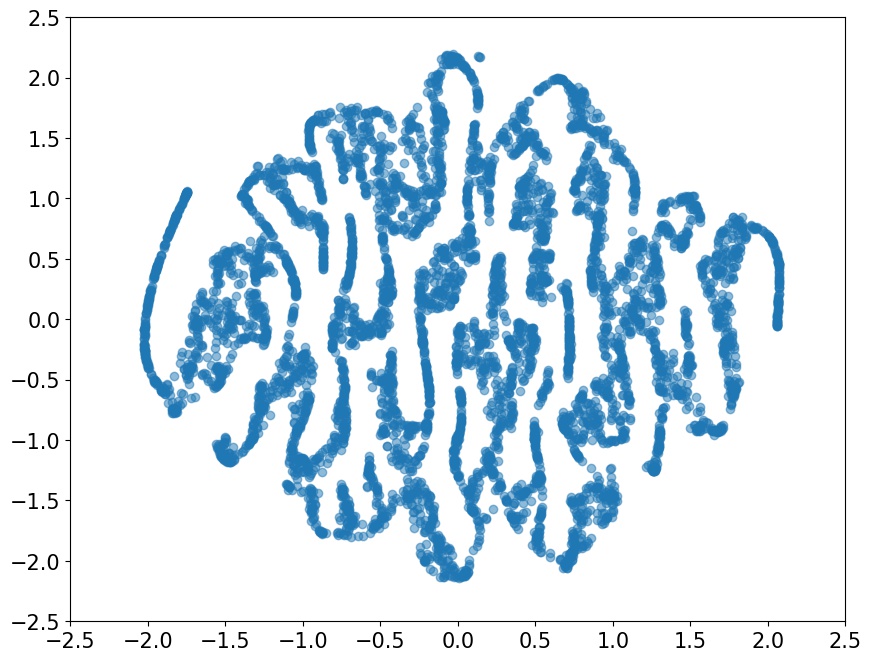}
}
\subfigure[The embeddings of item 482 on the Movielens1M dataset.]{
\includegraphics[width=0.2381\linewidth]{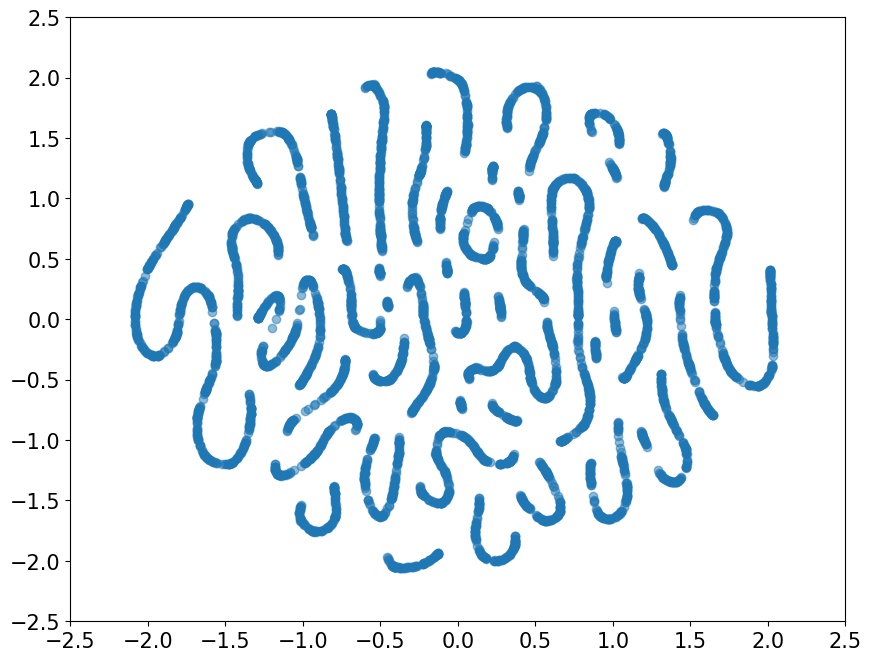}
}
\subfigure[The weights of the hidden layer on the Ciao dataset.]{
\includegraphics[width=0.2381\linewidth]{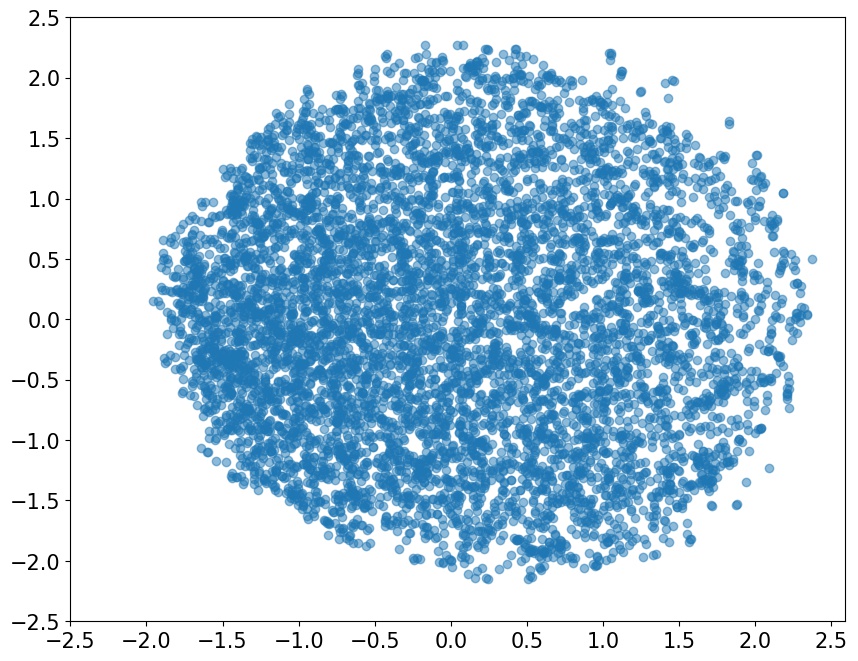}
}
\subfigure[The embeddings of item 8271 on the Ciao dataset.]{
\includegraphics[width=0.2381\linewidth]{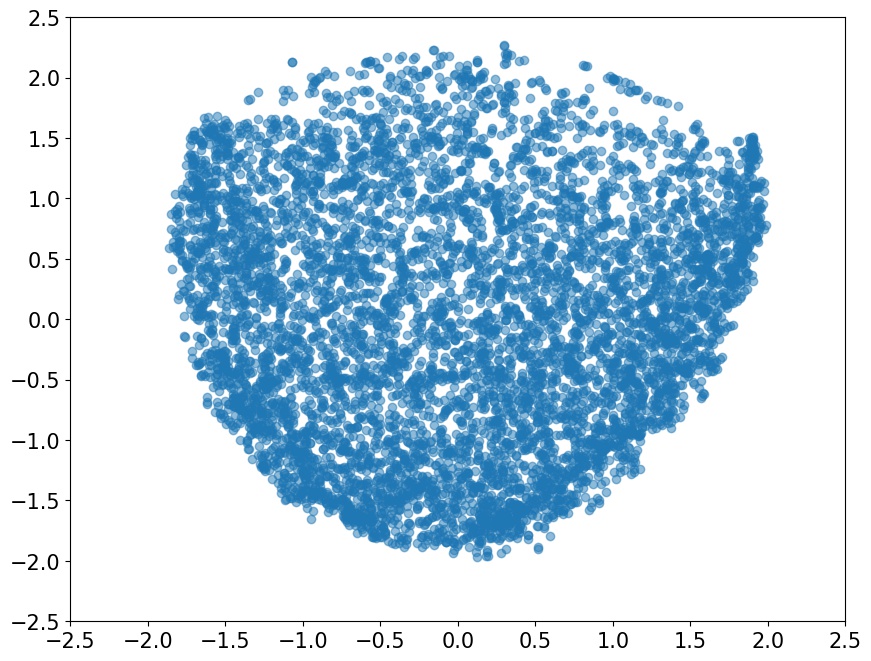}
}
\caption{The generated weights and item embeddings reduced dimension by t-SNE and normalized by mean and standard deviation on the four datasets, where one point corresponds to one user.}
\label{fig_5_1}
\end{figure*}

Next, we want to understand whether \ac{MetaMF} shows a high capacity at a small \ac{RP} model scale compared to \ac{NCF}.
And to which degree does \ac{MetaMF} generate different item embeddings and \ac{RP} models for different users while exploiting collaborative filtering?

\subsection{Model Scale Analysis}
\label{RQ3}

We examine whether \ac{MetaMF} shows a high model capacity at a small \ac{RP} model scale by comparing \ac{MetaMF} with \ac{NCF} at different model scales on the Douban and Hetrec-movielens datasets.
Because \ac{MetaMF} and \ac{NCF} are both \ac{MLP}-based methods, we represent each model scale as a combination of the item embedding size and the list of layer sizes,\footnote{The specific number of parameters is $d_u\times{m}+d_i\times{n}+\sum_{l=1}^{L}f^l_{in}\times{f^l_{out}}$, here we use $l$ to differentiate different layers.} which are the key hyper-parameters to affect the number of parameters.
For \ac{MetaMF}, we also list the collaborative vector size and the hidden layer size in the \ac{CM} and \ac{MR} modules for each model scale, however we only care about the generated parameters in the prediction module of \ac{MetaMF} because in federated recommender systems we only need to deploy the prediction module (i.e., the \ac{RP} model) on local devices.
Note that the number of parameters we consider is independent from the number of users/devices.

In Table~\ref{table_5_3} and Table~\ref{table_5_4}, we list the performance of \ac{MetaMF} and \ac{NCF} for different model scales.
By comparing the best settings of \ac{MetaMF} $(32,[8],[128,512])$ and \ac{NCF} $(256,[256,128,64])$, we see that \ac{MetaMF} achieves a comparable performance with \ac{NCF} with a smaller item embedding size, fewer layers and smaller layer sizes.
Importantly, at small model scales, \ac{MetaMF} significantly outperforms \ac{NCF}.
This is because the item embeddings and the \ac{RP} model generated by \ac{MetaMF} are private, using a small scale is sufficient to accurately encode the preference of a specific user and predict her ratings.
However for \ac{NCF}, the shared \ac{RP} model with item embeddings has to be large in size since it needs to incorporate the information of all users and predict ratings for all users.
We conclude that generating private item embeddings and \ac{RP} models helps \ac{MetaMF} to keep a relatively high capacity with fewer parameters in item embeddings and \ac{RP} models, and improve the performance of rating prediction at the small model scale.
Furthermore, when deploying the recommendation system on mobile devices, this advantage of \ac{MetaMF} can save storage space, RAM, energy and communication bandwidth.

As the model scale increases, the performance of \ac{MetaMF} deteriorates earlier than \ac{NCF}.
Because there is not sufficient private data to train too many parameters for each user, larger model scales easily lead \ac{MetaMF} to overfit.
And even though we have the \ac{RG} strategy to alleviate the memory and computational requirements for generating private item embeddings, it is unrealistic to generate larger embeddings for too many items. 
For example, we do not list the performance of \ac{MetaMF} with the model scale of $(64,[16],[256,1024])$ on the Douban dataset, because there are too many items in the dataset, making generation too difficult.

\subsection{Weights and Embeddings}
\label{RQ4}

In order to verify that \ac{MetaMF} generates private item embeddings and \ac{RP} models for users while efficiently exploiting \ac{CF}, we visualize the generated weights and item embeddings after reducing their dimension by t-SNE~\citep{maaten2008visualizing} and normalizing them by mean and standard deviation,\footnote{Here, $\mathrm{norm}(x)=\frac{x-\mu}{\sigma}$, where $\mu$ is the mean and $\sigma$ is the standard deviation.} where each point represents a user's weights or item embeddings.
Because there are many items, we randomly select one item from each dataset for visualization.
As shown in Fig.~\ref{fig_5_1}, \ac{MetaMF} generates different weights and item embeddings for different users on most datasets, which indicates that \ac{MetaMF} has the ability to capture private factors for users.
We also notice the existence of many non-trivial clusters in most visualizations, which shows that \ac{MetaMF} is able to share information among users to take advantage of collaborative filtering in the meta network.

The only exception is on the Ciao dataset, where \ac{MetaMF} seems unable to learn distinguishable weights and item embeddings.
The Ciao dataset does not provide sufficient private data for learning effective weights and item embeddings for each user.
It also illustrates why \ac{MetaMF} does not perform well on the Ciao dataset.


\section{Conclusion and Discussion}
\label{section:conclusion}
In this paper, we studied the federated rating prediction problem.
In particular, we investigated how to reduce the model scale of \acl{MF} methods in order to make them suitable for a federated environment.
To achieve this, we proposed a novel \acl{MF} framework, named \ac{MetaMF}, that can generate private \ac{RP} models as well as item embeddings for each user with a meta network.
We conducted extensive experiments to compare and analyze the performance of \ac{MetaMF}.
\ac{MetaMF} performs competitively, at the level of state-of-the-art \ac{RP} methods despite using a significant smaller \ac{RP} model and embedding size for items.
In particular, by using \acl{CF} in a federated environment, \ac{MetaMF} outperforms the federated recommendation method FedRec by a large margin.
Thus, we hope \ac{MetaMF} can advance future research on federated recommendation by presenting a new framework and a scheme.

Next, we discuss some limitations of \ac{MetaMF} and related future work.
First, \ac{MetaMF} still has a risk of leaking private information.
\ac{MetaMF} uses a meta network to directly generate private item embeddings and \ac{RP} models on the server. 
Although the meta network can efficiently exploit \ac{CF} to improve \ac{RP} performance, it may leak personal information about users in private item embeddings, \ac{RP} models and their updates.
As future work, we plan to design a more privacy-aware generation network that preserves the high \ac{RP} performance at the same time.
Second, currently \ac{MetaMF} cannot handle cold-start users well~\cite{bobadilla2012collaborative}.
Although \ac{MetaMF} incorporates a \ac{CM} module to collect collaborative information to alleviate this issue, it still needs a certain amount of data for each user to achieve satisfactory performance.
As a result, it does not perform well when there is not enough personalized data for each user.
A possible solution direction is to reduce the data requirements of \ac{MetaMF} using few-shot~\cite{DBLP:journals/corr/abs-1904-05046} or zero-shot learning~\cite{DBLP:journals/pami/XianLSA19}.
Finally, because the ranking prediction task~\cite{li2017narm} is also important in the area of recommendation system, we will also evaluate the performance of \ac{MetaMF} on the ranking prediction. 


\section*{Data and Code}
To facilitate reproducibility of our work, we are sharing all resources used in this paper at 
\url{https://github.com/TempSDU/MetaMF}. 

\begin{acks}
We thank our anonymous reviewers for their helpful comments. 

This work is supported by National Key R\&D Program of China with grant No. 2019YFB2102600, the Natural Science Foundation of China (61832012, 61972234, 61902219), the Foundation of State Key Laboratory of Cognitive Intelligence, iFLYTEK, P.R. China (COGOSC-20190003), the Key Scientific and Technological Innovation Program of Shandong Province (2019JZZY010129), the Fundamental Research Funds of Shandong University, and the Innovation Center for Artificial Intelligence (ICAI).

All content represents the opinion of the authors, which is not necessarily shared or endorsed by their respective employers and/or sponsors.
\end{acks}

\bibliographystyle{ACM-Reference-Format}
\bibliography{references}

\end{document}